\documentclass[10pt, conference, compsocconf]{IEEEtran}
\usepackage{cite}
\usepackage{url}

\ifCLASSINFOpdf
\else
  \usepackage[dvips]{graphicx}
   \graphicspath{{../eps/}}
   \DeclareGraphicsExtensions{.eps}
\fi

\begin{document}
\title{Application-Network Collaboration Using SDN for Ultra-Low Delay Teleorchestras}

\author{Emmanouil Lakiotakis\IEEEauthorrefmark{1}\IEEEauthorrefmark{2}, Christos Liaskos\IEEEauthorrefmark{1},
Xenofontas Dimitropoulos\IEEEauthorrefmark{1}\IEEEauthorrefmark{2}\\
{\small{}\IEEEauthorrefmark{1}Foundation for Research and Technology
- Hellas (FORTH)}\\
{\small{}\IEEEauthorrefmark{2}University of Crete, Computer Science
Department}\\
{\small{}Emails: \{manoslak,cliaskos,fontas\}@ics.forth.gr}\\

}

\maketitle

\begin{abstract}
Networked Music Performance (NMP) constitutes a class of ultra-low delay sensitive applications, allowing geographically separate musicians to perform seamlessly as a tele-orchestra. For this application type, the QoS indicator is the mouth-to-ear delay, which should be kept under $25$ milliseconds. The mouth-to-ear delay comprises signal processing latency and network delay. We propose a strong collaboration between the network and NMP applications to \emph{actively} keep the to mouth-to-ear delay minimal, using direct state notifications. Related approaches can be characterized as \emph{passive}, since they try to estimate the network state indirectly, based on the end application performance.  Our solution employs Software Defined Networking (SDN) to implement the network-to-application collaboration, being facilitated by the well-defined network interface that SDN offers. Emulation results show that the proposed scheme achieves an improvement of up to $59\%$ in mouth-to-ear delay over the existing passive solutions.

\end{abstract}


\IEEEpeerreviewmaketitle

\section{Introduction}

Many applications used in daily life require responsive Internet connectivity. In this category
belong instant messaging services, social network services, world wide web browsing, multimedia streaming, financial transactions etc. Among the above cases, there are
subcategories that are much more demanding in their QoS restrictions than others. For instance, multimedia
streaming services require low-latency connectivity. In the present paper,
we conduct a study of the interaction between an application and the network, focusing on Networked Music Performance (NMP) systems.

The term NMP was initiated by John Lazzaro from Berkeley University in 2001 and since then the term is globally used
for describing real time distant musical interaction using the Internet~\cite{lazzaro2001case}.
NMP describes the process where musicians located in different places around the world perform together via the Internet.
This process has very low delay tolerance. More specifically, in NMP services the maximum affordable delay between
the transmitted and the finally played signal should be up to 25 ms. This constraint is denoted as Ensemble Performance Threshold (EPT)~\cite{schuett2002effects}.
Thus, NMP systems can be approached as QoS-sensitive applications, whose evaluation metric is the end-to-end delay.

The present work studies NMP performance from the aspect of direct collaboration between such a system and the network.
Differentiating from the existing approaches, we consider a case where the network can directly inform an NMP system of its current or expected status (e.g., incoming traffic congestion).
NMPs can then alter their signal processing parameters, keeping the end-to-end delay under the EPT threshold.
The proposed approach is implemented and evaluated in a realistic, emulated setup.
The SDN technology is used towards this end, given the inherent ease in interacting with the network as a whole via a controller.

The remainder of the paper is organized as follows. Section \ref{endtoendanalysis} provides the necessary background and details  of the employed system model. Section \ref{related} presents the related work in this problem domain. Section \ref{methodology} introduces our architecture. Evaluation is discussed in Section \ref{results} and finally, Section \ref{conclusion} concludes our work.

\section{Background}
\label{endtoendanalysis}
Two facts affect the performance of NMP systems: the first factor refers to the delay related to the audio context. In this aspect, the delay is caused by the signal capturing from the audio hardware, the audio coding in the transmitter's side and decoding in the receiver's side. The second factor refers to the delay caused by the transmission of data via the network equipment.

Regarding the network delay, in an ideal scenario, routers should forward the packets that they receive instantly but in cases of bandwidth overload this is not feasible. This also explains the jitter that appears and affects data transmissions. Delay due to queuing means that the total network delay is higher than the physical distance between peers. Apart from the delay caused by routing policies, delay is also caused by limitations in bandwidth offered to users by Internet Service Providers (ISPs). Conventional Internet connections, such as DSL, make NMP impossible, since even a small ICMP (Internet Control Message Protocol) packet has response time over 50 ms which is twice the value of EPT as mentioned above. Using audio compression techniques would be an important solution towards reducing bit-rate to required levels but conventional audio coders increase latency due to encoding/decoding process and this is not acceptable in NMPs. For instance, standard coders like MP3 or AAC have a delay of about 100 ms or more. Even AAC-Low Delay encoder still introduces delay of about 20 ms using 48 kHz sampling rate \cite{goto1910virtual}. This prevents the use of conventional audio encoding/decoding methods in NMP in the general case.

\begin{figure}[!t]
\centering
\includegraphics[width=3.5in]{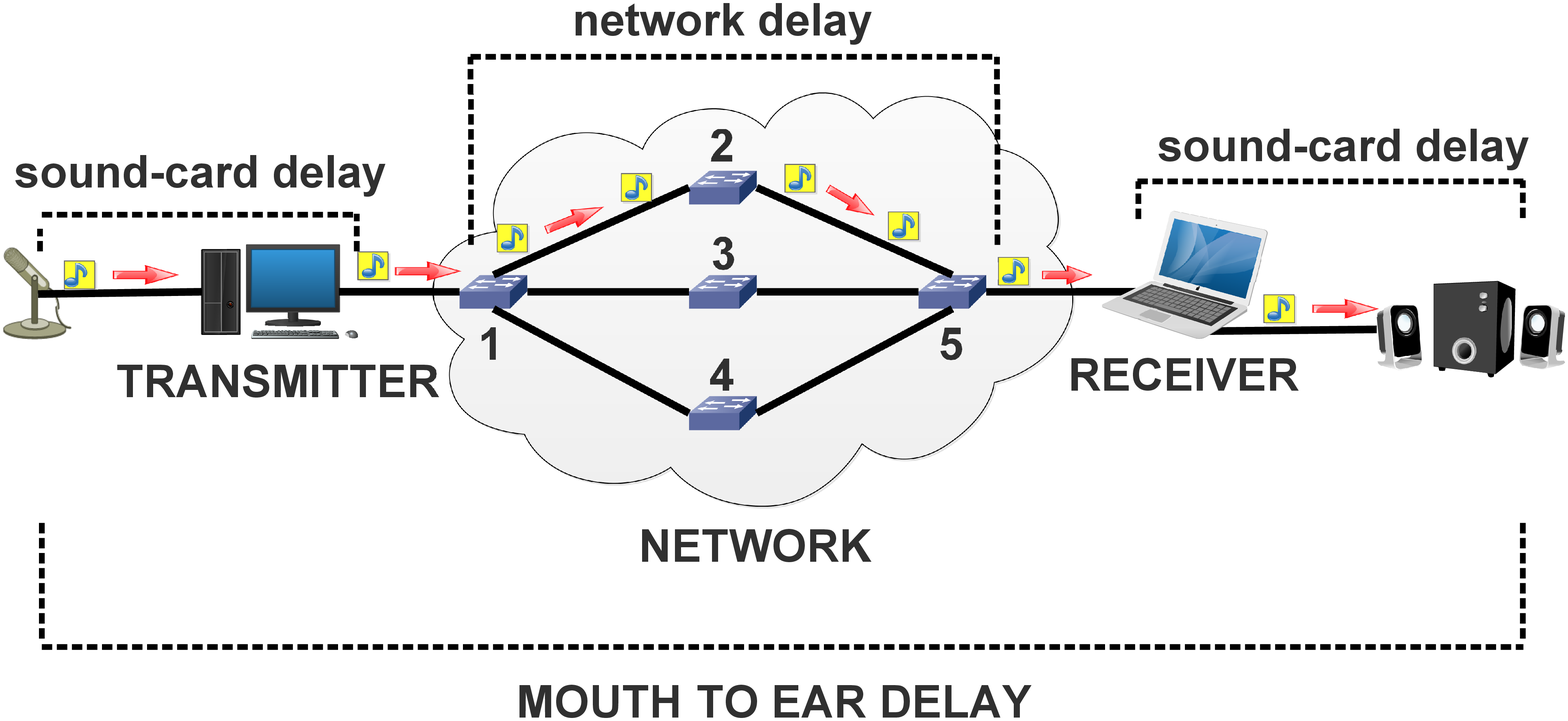}
\caption{End-to-end delay in NMP systems}
\label{end_to_end_delay}
\end{figure}

The overall delay for an audio signal to propagate from the transmitter's mouth to the receiver's ear is called \emph{mouth-to-ear} delay. This delay, depicted in Fig.~\ref{end_to_end_delay}, can be expressed as follows:
\begin{equation} \label{eq:1}
d_{mouth-to-ear}=d_{a-s} +d_{p-s} + d_{n} + d_{a-d} +d_{p-d}
\end{equation}
where $d_{mouth-to-ear}$ denotes mouth-to-ear delay, $d_{a-s}$ is the delay inserted by the transmitter's sound-card, $d_{n}$ is the delay added due to transmission through the network and $d_{a-d}$ is the delay inserted by the receiver's sound-card. $d_{p-s}$ and $d_{p-d}$ describe the delay inserted due to audio processing and encoding/decoding in transmitter/receiver side. In this paper uncompressed audio is transmitted so equation (\ref{eq:1}) is transformed to equation (\ref{eq:2}).
\begin{equation} \label{eq:2}
d_{mouth-to-ear}=d_{a-s} + d_{n} + d_{a-d}
\end{equation}

In cases that transmitter-receiver use sound-cards with similar specifications regarding to reading/recording processes which means that $d_{a-s}$ = $d_{a-d}$ = $d_{sound-card}$, Equation (\ref{eq:2}) evaluates end-to-end delay in NMP systems.
\begin{equation}
d_{mouth-to-ear}=2*d_{sound-card} + d_{n}
\label{eq:3}
\end{equation}


Equation (\ref{eq:3}) models the mouth-to-ear delay as a function of delay created by audio capturing in transmitter/receiver side and delay caused by transmission through the network. In the audio community, the delay caused by audio capturing is called blocking delay. It indicates the delay due to processing by sound-card. Equation (\ref{eq:4}) describes the blocking delay evaluation process.
\begin{equation}
d_{blocking-delay}= \frac{frame\ size}{sampling\ rate} + d_{0}
\label{eq:4}
\end{equation}

In equation (\ref{eq:4}) frame size denotes the size of audio packets that sound-card can process per hardware clock tick and the sampling rate is the number of samples the sound-card acquires per second. Finally, $d_{0}$ is a constant delay that is due to the sound-card's hardware quality. It is obvious that to achieve blocking delay minimization, the fraction between frame size and sampling rate should be minimized.

\section{Related Work}
\label{related}
This section first gives an overview of previous research on NMP systems and SDN-supported mechanisms for QoS-aware applications.
Research has approached NMP systems from two different perspectives: audio processing and network. Both approaches have a common feature: approaching the problem from a single perspective (audio-latency or network-latency only) does not allow important improvement because NMP is a summary of both perspectives.

From the audio perspective, many researchers focus on the audio flows forwarding process. In more details, not all participants are interested in receiving audio from all transmitters. For this reason, participants should declare their interests and forwarding is based on this profiling type. This method is implemented by Selecting Forwarding Unit (SFU) \cite{baltas_ultra_2014,carot2007network,carot2006network,carot2007networked,xiaoyuan_gu_network_centric_2005}. On the other hand, collecting, mixing and forwarding all audio flows is proposed using another entity called Multipoint Conferencing Unit (MCU) \cite{akoumianakis_musinet_2014,xylomenos2013reduced}. Additionally, a common trend in NMP projects is that Session Initiation Protocol (SIP) is used to support control messages among transmitter and receiver side. SIP is a protocol widely used in parallel with Real Time Protocol (RTP) for initial handshaking and dynamic transmission modifications during runtime \cite{nam_towards_2014,nurmela2007session,ali2013session,sinnreich_internet_2006,camarillo_evaluation_2003}.

From the network perspective, SDN is widely used in network condition-aware applications~\cite{Liaskos1,Liaskos2,Liaskos3,Liaskos4}. The main feature is that traffic is prioritized using criteria such as Type of Service (TOS), requirements, SLAs or packet header fields~\cite{sieber_network_2015,egilmez_distributed_2014,adami_network_2015,sharma_demonstrating_2014}. Moreover, SIP combined with SDN is introduced also in \cite{maribondo_avoiding_2016} where an approach for VoIP applications is described by codec modification due to network changes. Finally, in~\cite{tomovic_sdn_2014} an alternative method for collecting network statistics is introduced where SDN Controller sends periodically requests to switches about statistics. This information is used for network delay monitoring by SDN Controller.

All solutions described above examine NMP either from signal processing or network perspective but they do not take into account both delay types. Approaching NMP from the audio perspective leads in innovating in audio encoding/decoding methods that reduce blocking delay. Additionally, selective audio forwarding through the network can contribute in traffic congestion cases but this inserts additional delay caused by the pre-processing stage for filtering audio flows. On the other hand, from network perspective, exploiting SDN capability of global network view and dynamic adaptation to network changes allows optimal path selection for audio transmission but it ignores blocking delay that comes into play during NMP process.

In our proposed architecture, we approach end-to-end delay in NMP systems combining the two individual perspectives, i.e., the audio processing and the network delay. In more details, during NMP operation, the two basic components that participate in a NMP system, application and network, can interact in order to overcome network delay increase and keep end-to-end delay constant despite traffic congestion problem. This is achievable by modifying audio process that results in blocking delay decrease. In other words, network delay increases can be absorbed by blocking delay decreases, offering seamless quality of service.

\section{Proposed Approach}
\label{methodology}

\begin{figure}[!t]
\centering
\includegraphics[width=3.5in]{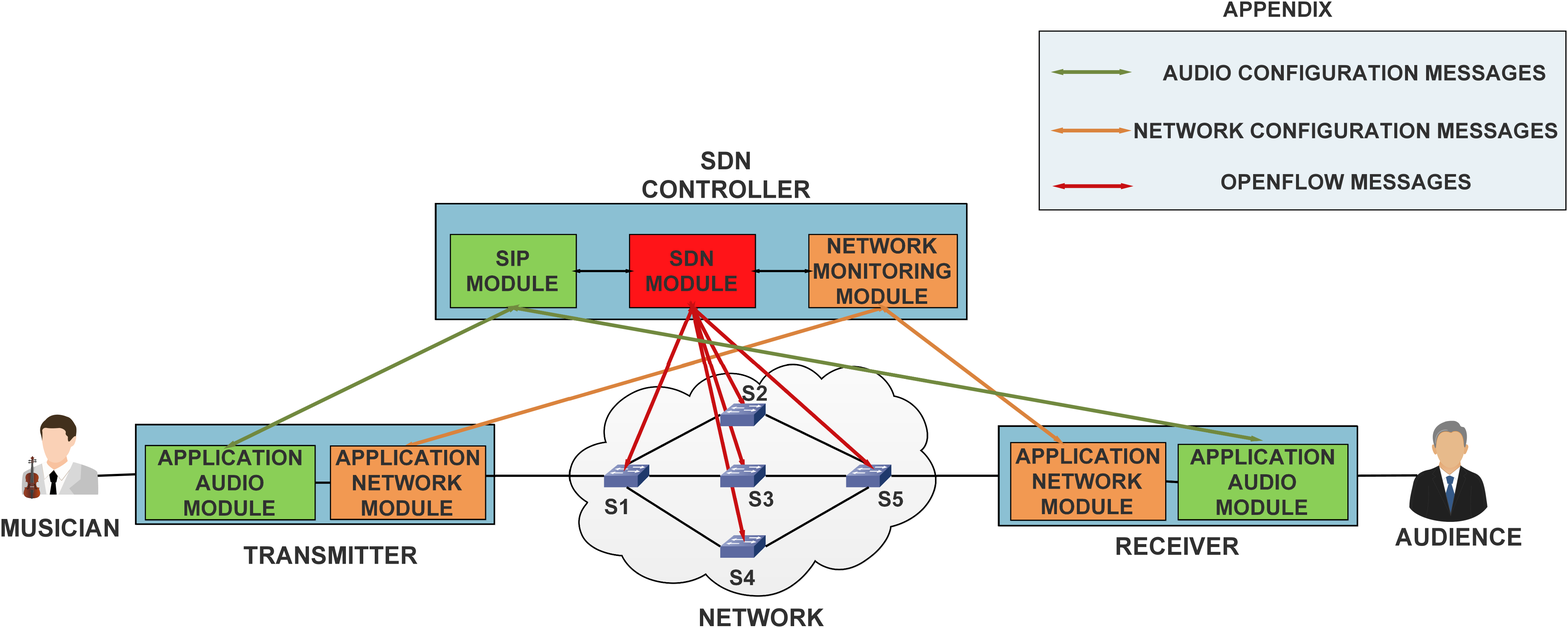}
\caption{Proposed architecture}
\label{abstract_figure}
\end{figure}

In our implementation, depicted in Fig.~\ref{abstract_figure}, SDN is used to increase NMP performance during link congestion. End-to-end delay is monitored in real time and rerouting decisions are taken in cases that another path shows less network delay according to a threshold value. In case that all paths are congested, (which can lead to over-EPT end-to-end delay), a request is sent to the application to modify its audio processing configuration, reducing the blocking delay, coping with the network delay increase. During the NMP process there are three key roles: transmitter, receiver and SDN Controller. Each of the above entities are equipped with modules that implement the interaction between network and application.
\subsection{Transmitter}
In the proposed architecture, the Transmitter component is any entity that generates audio. For instance, transmitters are musicians that participate in the teleorchestra application. The Transmitter component is equipped with two modules: the Application Audio module and the Application Network module. The Application Audio module informs the SDN Controller about the audio profile of transmitter and also captures the audio signal for transmission through the network. Audio profiling describes the initialization process (occurring once, during system setup) where each participant tests his sound-card performance for various frame size and sampling rate combinations. Thus, a dictionary of signal configuration to blocking delay is created. This information forms the audio profile of each user and is sent to SDN Controller in order to have a summary of audio performance for each user. The Application Audio module also receives requests for audio processing modification in case of traffic congestion. This operation is similar to the Session Initiation Protocol (SIP), where all users negotiate to the audio transmission parameters~\cite{sip}.

The second module in the transmitter's side is the Application Network module. This module transmits audio signal through the network. Also, the Application Network module is responsible for network delay monitoring. Each transmitter records in real time the network delay towards the corresponding receiver and informs the SDN Controller. Following this process, the SDN Controller uses the collected information to estimate the mouth-to-ear delay using equation (\ref{eq:3}).

\subsection{Receiver}
The Receiver entity represents all users in the NMP system that receive audio flows. In this category belong the audience users and also musicians that should receive audio flows from other musicians in order to be synchronized. The receiver component has the same modules as the transmitter but they are adapted to its role. In more details, the receiver uses the Application Audio module for informing the SDN Controller about its audio profile and playing the received audio signal. The Application Audio module receives requests from the SDN Controller in case of traffic congestion, in order to change its audio processing configuration. In addition to the Application Audio module, the receiver uses the Application Network module in order to receive the audio signal from network and participate in the traffic monitoring process, described later.
\subsection{SDN Controller}
The SDN Controller is the major component of our architecture. It combines the conventional SDN Controller duties for taking routing decisions based on the network and application performance. It is also assigned to interact with applications, informing them on traffic congestion, allowing them to modify the audio processing parameters, coping with the network delay increase.

The requested SDN Controller functionality is implemented by three modules:
The SIP module is responsible for collecting audio profiles from each user when he joins into the application. For this reason, it keeps a data structure that stores the audio profile information for each user. It also informs the application when the network is congested in order to choose another frame size and sampling rate combination, thus decreasing the blocking delay. The second module that is used in the SDN Controller entity is the SDN module. This module is responsible for installing flow rules into network switches that participate into the selected path. The communication between the SDN module and the switches is accomplished via the OpenFlow protocol~\cite{akyildiz_roadmap_2014}. Finally, the third module that SDN Controller uses is the Network Monitoring module. This module is responsible for monitoring the network delay. It keeps real-time measurements of network delay per each path in the network in a data structure. This information is used in case that a rerouting decision is required to deal with a traffic congestion problem.

\subsection{Network delay monitoring}
\label{network_monitoring}
As we describe in Section~\ref{methodology}, the Network Monitoring module in the SDN Controller is used for monitoring the network delay for each path in the network. Many approaches that use network monitoring (mentioned in Section \ref{related}), measure connection events at the switches (such as PACKET\_INs).

\begin{figure}[!t]
\centering
\includegraphics[width=2.5in]{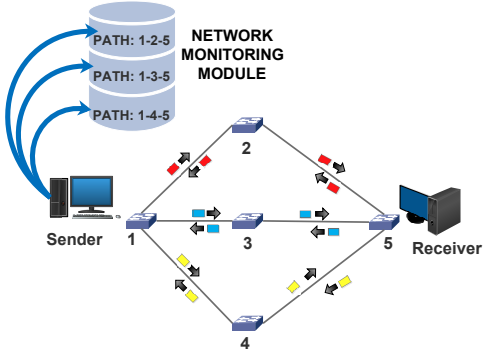}
\caption{Network delay monitoring process}
\label{ping_process}
\end{figure}

In our approach, switches do not participate in network monitoring process, reducing their load. This operation is moved to the end hosts that communicate through the network. The  process is depicted in Fig.~\ref{ping_process}. A sender node sends periodically UDP packets, with specific header fields, over each path towards the receiver. The receiver acts as an echo server and sends these packets back over the paths that were initially used. The sender then evaluates the network delay in terms of Round Trip Time (RTT), and the results are sent to the Network Monitoring module.
\subsection{Rerouting process}
\label{rerouting_process}
Following the network monitoring process described in \ref{network_monitoring}, the SDN Controller has a view of the network delay for each path in the network in real time. The SDN Controller, via the Network Monitoring module, can choose the path that results in below-EPT mouth-to-ear delay, given the current blocking delay. In case that a path yields a network delay less than a threshold value, the SDN module chooses this path and installs the appropriate flow rules to the switches that form it, in an application-transparent manner.
\subsection{Application-Network interaction}
In subsection \ref{rerouting_process}, we describe the process by which the SDN Controller evaluates the mouth-to-ear delay, using equation (\ref{eq:3}). In case that there is at least one path that results in below-EPT end-to-end delay, the SDN module chooses it. On the other hand, if no such a path exists, the SDN Controller informs the application to modify its audio processing configurations at both the transmitter and receiver-side. This interaction aims to cope with the network delay increase via incurring a corresponding blocking delay decrease. For the communication between the application and network, the SIP module in the SDN Controller sends notifications to the Application Audio module at both the transmitter and the receiver-side. The modification of audio processing parameters may cause a drop in the sound quality. Care is taken to ensure the use of configurations that offer acceptable quality, while keeping the  mouth-to-ear delay under the EPT value.
\section{Evaluation}
\label{results}
We employed an emulation scenario to evaluate the efficiency of an NMP process running over the proposed network-to-application interaction system. The goal is to demonstrate how application-network collaboration can improve application robustness against sudden network changes.
\subsection{Emulation scenario}
\begin{figure}[!t]
\centering
\includegraphics[width=3.5in]{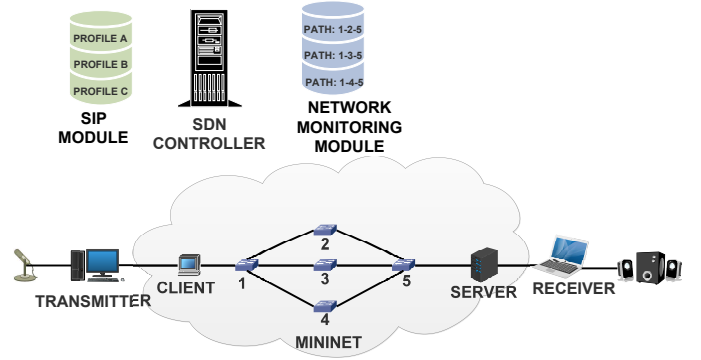}
\caption{Emulation scenario}
\label{emulated_scenario}
\end{figure}

The emulation scenario is developed in Mininet~\cite{sharma2014mininet}. As shown in Fig.~\ref{emulated_scenario}, it consists of five OpenFlow switches~\cite{akyildiz_roadmap_2014}, that form the paths between the transmitter and the receiver. We employ the POX SDN Controller, which is widely used in SDN research~\cite{kaur2014network}. Audio processing and streaming functionality is implemented using Mathworks Simulink environment at both the transmitter and the receiver sides~\cite{MATLAB:2010}. Finally, congestion is emulated using the Netem traffic control tool~\cite{tc_tool}. The default audio configuration set refers to 22050 Hz as sampling rate and 128 samples frame size and the alternative audio processing set refers to 44100 Hz sampling rate and 64 samples frame size.
\subsection{Emulation results}
\begin{figure}[!t]
\centering
\includegraphics[width=3.7in]{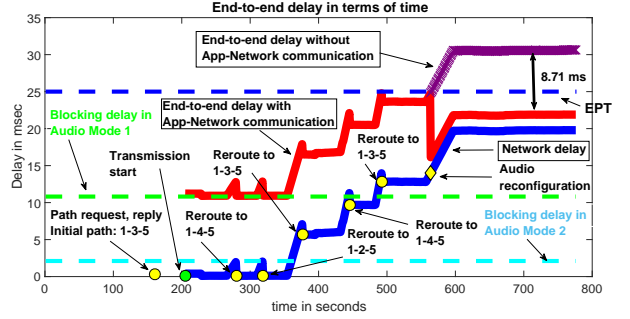}
\caption{End-to-end delay, network delay and blocking delay}
\label{evaluated_end}
\end{figure}

In order to test the performance of our architecture, we introduce latency increases at the paths of Fig.~\ref{emulated_scenario} sequentially. Figure~\ref{evaluated_end} describes the resulting mouth-to-ear delay, network delay and blocking delay as a function of time. All rerouting events are depicted as circles in the Figure.

In the described experiment, initially the SDN Controller assigns the fastest path for audio transmission. This path assignment takes place at t=161 s, selecting path $1-3-5$ based on the minimum network delay. At t=200 s, the Networked Music Performance starts using this selected path for audio transmission. We used NETEM in order to add delay to the path and we tested the rerouting process. By increasing the delay of path $1-3-5$, at t=280 s the SDN Controller reroutes audio flows to path $1-4-5$, as the difference between its delay and initially selected path is greater than 2 ms, which is selected as a threshold value for rerouting decisions. The same process is repeated, adding delay to path $1-4-5$ with NETEM, resulting into another rerouting of application traffic to path $1-2-5$. Table~\ref{transition_array} describes the ensuing rerouting events and audio modifications that took place.

Finally, when the network delay increase results in over-EPT mouth-to-ear delay, the SIP module informs the application side for audio modification. In this example, audio modification decision is taken at t=564 s and instantly the application switches from the default to the alternative audio configuration set which introduces less blocking delay. This results in decreasing the end-to-end delay despite the network delay increase.

In Fig~\ref{evaluated_end} we also compare our approach with the case where the network does not inform the application for an audio modification. It is shown that the  interaction between the application and the network benefits the mouth-to-ear delay decrease by an average of 8.71 ms, without violating the EPT constraint. As it is also shown, without the proposed interaction, this would not be possible, resulting in over 30 ms mouth-to-ear delay, disrupting the NMP operation.
\begin{table}
\renewcommand{\arraystretch}{1.3}
\caption{Transition table}
\label{transition_array}
\centering
\begin{tabular}{|c||c||c||c|}
\hline
Time (s) & Current path & Next path & Action\\
\hline
161 & - & 1-3-5 & Path assignment\\
280 & 1-3-5 & 1-4-5 & Rerouting\\
319 & 1-4-5 & 1-2-5 & Rerouting\\
377 & 1-2-5 & 1-3-5 & Rerouting\\
446 & 1-3-5 & 1-4-5 & Rerouting\\
493 & 1-4-5 & 1-2-5 & Rerouting\\
564 & 1-2-5 & 1-2-5 & Audio modification\\
\hline
\end{tabular}
\end{table}
In order to quantify the degree that our system improved end-to-end delay, we defined the gain metric. Gain value is given from equation (\ref{gain}):
\begin{equation}
gain= \frac{d_{audio\ mode1}- d_{audio\ mode2}}{d_{audio\ mode1}} *100\%
\label{gain}
\end{equation}
where $d_{audio\ mode1}$ and $d_{audio\ mode2}$ denote the mouth-to-ear delay with and without audio modification. In the described process, the average gain was equal to 28.32\%. During the experimental evaluation, maximal results reach up to 59\% delay improvement. This value is a function of the sampling rate and the frame size that will be selected in order to decrease the blocking delay.

\section{Conclusion}
\label{conclusion}
This work introduced a novel framework for collaboration between application and network. The goal of the framework
is to cope with sudden network changes and traffic congestion problem offering guaranteed end-to-end delay. This objective was formulated in terms of the basic factors that affect delay in Networked Music Performance systems. The employed system model provided insights on real time traffic congestion detection and providing powerful solutions to its mitigation via application and network interaction. The same strategy can be applied in other extremely delay-sensitive application types, such as online gaming. The insights were validated realistically within an emulated, SDN setup.


\section*{Acknowledgment}
This work has been funded by the European Research Council Grant Agreement no. 338402, project NetVolution (\url{http://netvolution.eu}).

\bibliographystyle{IEEEtran}

\begin{thebibliography}{10}
\providecommand{\url}[1]{#1}
\csname url@samestyle\endcsname
\providecommand{\newblock}{\relax}
\providecommand{\bibinfo}[2]{#2}
\providecommand{\BIBentrySTDinterwordspacing}{\spaceskip=0pt\relax}
\providecommand{\BIBentryALTinterwordstretchfactor}{4}
\providecommand{\BIBentryALTinterwordspacing}{\spaceskip=\fontdimen2\font plus
\BIBentryALTinterwordstretchfactor\fontdimen3\font minus
  \fontdimen4\font\relax}
\providecommand{\BIBforeignlanguage}[2]{{%
\expandafter\ifx\csname l@#1\endcsname\relax
\typeout{** WARNING: IEEEtran.bst: No hyphenation pattern has been}%
\typeout{** loaded for the language `#1'. Using the pattern for}%
\typeout{** the default language instead.}%
\else
\language=\csname l@#1\endcsname
\fi
#2}}
\providecommand{\BIBdecl}{\relax}
\BIBdecl

\bibitem{lazzaro2001case}
J.~Lazzaro and J.~Wawrzynek, ``A case for network musical performance,'' in
  \emph{Proceedings of the 11th international workshop on Network and operating
  systems support for digital audio and video}.\hskip 1em plus 0.5em minus
  0.4em\relax ACM, 2001, pp. 157--166.

\bibitem{schuett2002effects}
N.~Schuett, ``The effects of latency on ensemble performance,'' \emph{Bachelor
  Thesis, CCRMA Department of Music, Stanford University}, 2002.

\bibitem{goto1910virtual}
M.~Goto, I.~Hidaka, H.~Matsumoto, Y.~Kuroda, and Y.~Muraoka, ``A virtual jazz
  session system: Virja session,'' \emph{Transactions of Information Processing
  Society of Japan}, vol.~21, 1910.

\bibitem{baltas_ultra_2014}
G.~Baltas and G.~Xylomenos, ``Ultra low delay switching for networked music
  performance.''\hskip 1em plus 0.5em minus 0.4em\relax IEEE, Jul. 2014, pp.
  70--74.

\bibitem{carot2007network}
A.~Car{\^o}t and C.~Werner, ``Network music performance-problems, approaches
  and perspectives,'' in \emph{Proceedings of the “Music in the Global
  Village”-Conference, Budapest, Hungary}, vol. 162, 2007, pp. 23--10.

\bibitem{carot2006network}
A.~Car{\^o}t, U.~Kr{\"a}mer, and G.~Schuller, ``Network music performance (nmp)
  in narrow band networks,'' in \emph{Audio Engineering Society Convention
  120}.\hskip 1em plus 0.5em minus 0.4em\relax Audio Engineering Society, 2006.

\bibitem{carot2007networked}
A.~Car{\^o}t, P.~Rebelo, and A.~Renaud, ``Networked music performance: State of
  the art,'' in \emph{Audio engineering society conference: 30th international
  conference: intelligent audio environments}.\hskip 1em plus 0.5em minus
  0.4em\relax Audio Engineering Society, 2007.

\bibitem{xiaoyuan_gu_network_centric_2005}
{Xiaoyuan Gu}, M.~Dick, Z.~Kurtisi, U.~Noyer, and L.~Wolf, ``Network-centric
  music performance: practice and experiments,'' \emph{IEEE Communications
  Magazine}, vol.~43, no.~6, pp. 86--93, Jun. 2005.

\bibitem{akoumianakis_musinet_2014}
D.~Akoumianakis, C.~Alexandraki, V.~Alexiou, C.~Anagnostopoulou,
  A.~Eleftheriadis, V.~Lalioti, A.~Mouchtaris, D.~Pavlidi, G.~C. Polyzos,
  P.~Tsakalides, G.~Xylomenos, and P.~Zervas, ``The {MusiNet} project:
  {Towards} unraveling the full potential of {Networked} {Music} {Performance}
  systems.''\hskip 1em plus 0.5em minus 0.4em\relax IEEE, Jul. 2014, pp. 1--6.

\bibitem{xylomenos2013reduced}
G.~Xylomenos, C.~Tsilopoulos, Y.~Thomas, and G.~C. Polyzos, ``Reduced switching
  delay for networked music performance,'' in \emph{Packet Video Workshop
  (Poster Session)}, 2013.

\bibitem{nam_towards_2014}
H.~Nam, K.-H. Kim, J.~Y. Kim, and H.~Schulzrinne, ``Towards {QoE}-aware video
  streaming using {SDN}.''\hskip 1em plus 0.5em minus 0.4em\relax IEEE, Dec.
  2014, pp. 1317--1322.

\bibitem{nurmela2007session}
T.~Nurmela, ``Session initiation protocol,'' in \emph{Seminar on Transport of
  multimedia streams, University of Helsinki}.\hskip 1em plus 0.5em minus
  0.4em\relax Citeseer, 2007.

\bibitem{ali2013session}
A.~Ali, N.~Ahmad, M.~S. Akhtar, and A.~Srivastava, ``Session initiation
  protocol,'' \emph{International Journal of Scientific and Engineering
  Research}, vol.~4, no.~1, pp. 1--6, 2013.

\bibitem{sinnreich_internet_2006}
H.~Sinnreich and A.~B. Johnston, \emph{Internet communications using {SIP}:
  delivering {VoIP} and multimedia services with {Session} {Initiation}
  {Protocol}}, 2nd~ed.\hskip 1em plus 0.5em minus 0.4em\relax Indianapolis, IN:
  Wiley Pub, 2006, oCLC: ocm65340953.

\bibitem{camarillo_evaluation_2003}
G.~Camarillo, R.~Kantola, and H.~Schulzrinne,
  ``\BIBforeignlanguage{en}{Evaluation of transport protocols for the session
  initiation protocol},'' \emph{\BIBforeignlanguage{en}{IEEE Network}},
  vol.~17, no.~5, pp. 40--46, Sep. 2003.

\bibitem{Liaskos1}
C.~Liaskos, X.~Dimitropoulos, and L.~Tassiulas, ``Backpressure on the backbone:
  A lightweight, non-intrusive traffic engineering approach,'' \emph{IEEE
  Transactions on Network and Service Management}, vol. to appear, pp. 1--14,
  2016.

\bibitem{Liaskos2}
D.~Gkounis, V.~Kotronis, C.~Liaskos, and X.~Dimitropoulos, ``On the interplay
  of link-flooding attacks and traffic engineering,'' \emph{ACM SIGCOMM
  Computer Communication Review}, vol.~46, no.~1, pp. 5--11, 2016.

\bibitem{Liaskos3}
C.~Liaskos, V.~Kotronis, and X.~Dimitropoulos, ``A novel framework for modeling
  and mitigating distributed link flooding attacks,'' in \emph{IEEE
  INFOCOM'16}, 2016, pp. 1--9.

\bibitem{Liaskos4}
C.~Liaskos, ``A lightweight, non-intrusive approach for orchestrating
  autonomously-managed network elements,'' in \emph{IEEE ISCC'15}.\hskip 1em
  plus 0.5em minus 0.4em\relax IEEE, 2015, pp. 335--340.

\bibitem{sieber_network_2015}
C.~Sieber, A.~Blenk, D.~Hock, M.~Scheib, T.~Hohn, S.~Kohler, and W.~Kellerer,
  ``Network configuration with quality of service abstractions for {SDN} and
  legacy networks.''\hskip 1em plus 0.5em minus 0.4em\relax IEEE, May 2015, pp.
  1135--1136.

\bibitem{egilmez_distributed_2014}
H.~E. Egilmez and A.~M. Tekalp, ``Distributed {QoS} {Architectures} for
  {Multimedia} {Streaming} {Over} {Software} {Defined} {Networks},'' \emph{IEEE
  Transactions on Multimedia}, vol.~16, no.~6, pp. 1597--1609, Oct. 2014.

\bibitem{adami_network_2015}
D.~Adami, L.~Donatini, S.~Giordano, and M.~Pagano, ``A network control
  application enabling {Software}-{Defined} {Quality} of {Service}.''\hskip 1em
  plus 0.5em minus 0.4em\relax IEEE, Jun. 2015, pp. 6074--6079.

\bibitem{sharma_demonstrating_2014}
S.~Sharma, D.~Staessens, D.~Colle, D.~Palma, J.~Goncalves, M.~Pickavet,
  L.~Cordeiro, and P.~Demeester, ``Demonstrating resilient quality of service
  in {Software} {Defined} {Networking}.''\hskip 1em plus 0.5em minus
  0.4em\relax IEEE, Apr. 2014, pp. 133--134.

\bibitem{maribondo_avoiding_2016}
P.~D.~S. Maribondo and N.~C. Fernandes, ``\BIBforeignlanguage{en}{Avoiding
  {Voice} {Traffic} {Degradation} in {IP} {Enterprise} {Networks} {Using}
  {CAoS}}.''\hskip 1em plus 0.5em minus 0.4em\relax ACM Press, 2016, pp.
  34--36.

\bibitem{tomovic_sdn_2014}
S.~Tomovic, N.~Prasad, and I.~Radusinovic, ``{SDN} control framework for {QoS}
  provisioning.''\hskip 1em plus 0.5em minus 0.4em\relax IEEE, Nov. 2014, pp.
  111--114.

\bibitem{sip}
``Session initiation protocol, url =
  {http://www.networkworld.com/article/2332980/lan-wan/lan-wan-what-is-sip.html},.''

\bibitem{akyildiz_roadmap_2014}
I.~F. Akyildiz, A.~Lee, P.~Wang, M.~Luo, and W.~Chou,
  ``\BIBforeignlanguage{en}{A roadmap for traffic engineering in
  {SDN}-{OpenFlow} networks},'' \emph{\BIBforeignlanguage{en}{Computer
  Networks}}, vol.~71, pp. 1--30, Oct. 2014.

\bibitem{sharma2014mininet}
K.~K. Sharma and M.~Sood, ``Mininet as a container based emulator for software
  defined networks,'' \emph{International Journal of Advanced Research in
  Computer Science and Software Engineering}, vol.~4, no.~12, 2014.

\bibitem{kaur2014network}
S.~Kaur, J.~Singh, and N.~S. Ghumman, ``Network programmability using pox
  controller,'' in \emph{ICCCS International Conference on Communication,
  Computing \& Systems, IEEE}, no. s 134, 2014, p. 138.

\bibitem{MATLAB:2010}
``{MathWorks Simulink},''
  \url{https://www.mathworks.com/}.

\bibitem{tc_tool}
``{NETEM},'' \url{https://wiki.linuxfoundation.org/networking/netem}.

\end{thebibliography}

\end{document}